\documentclass[aps,prb,twocolumn,showpacs,groupedaddress]{revtex4}
\usepackage{amssymb}

\usepackage{graphicx}

\begin{document}

\title[]{Magnetoelectric Effect and Spontaneous Polarization in HoFe$_3$(BO$_3$)$_4$ and Ho$_{0.5}$Nd$_{0.5}$Fe$_3$(BO$_3$)$_4$}

\author{R. P. Chaudhury$^1$, F. Yen$^2$, B. Lorenz$^1$, Y. Y. Sun$^1$, L. N. Bezmaternykh$^3$, V. L. Temerov$^3$, and C. W. Chu$^{1,4,5}$}

\affiliation{$^1$ TCSUH and Department of Physics, University of Houston, Houston, Texas 77204-5002, USA}

\affiliation{$^2$ Applied Superconductivity Laboratory, Southwest Jiaotong University, Chengdu, Sichuan 610031, China}

\affiliation{$^3$ Institute of Physics, Siberian Division, Russian Academy of Sciences, Krasnoyarsk, 660036 Russia}

\affiliation{$^4$ Lawrence Berkeley National Laboratory, 1 Cyclotron Road, Berkeley, California 94720, USA}

\affiliation{$^5$ Hong Kong University of Science and Technology, Hong Kong, China}

\begin{abstract}
The thermodynamic, magnetic, dielectric, and magnetoelectric properties of HoFe$_3$(BO$_3$)$_4$ and Ho$_{0.5}$Nd$_{0.5}$Fe$_3$(BO$_3$)$_4$ are
investigated. Both compounds show a second order Ne\'{e}l transition above 30 K and a first order spin reorientation transition below 10 K.
HoFe$_3$(BO$_3$)$_4$ develops a spontaneous electrical polarization below the Ne\'{e}l temperature (T$_N$) which is diminished in external
magnetic fields. No magnetic-field induced increase of the polarization could be observed in HoFe$_3$(BO$_3$)$_4$. In contrast, the solid
solution Ho$_{0.5}$Nd$_{0.5}$Fe$_3$(BO$_3$)$_4$ exhibits both, a spontaneous polarization below T$_N$ and a positive magnetoelectric effect at
higher fields that extends to high temperatures. The superposition of spontaneous polarization, induced by the internal magnetic field in the
ordered state, and the magnetoelectric polarizations due to the external field results in a complex behavior of the total polarization measured
as a function of temperature and field.
\end{abstract}



\maketitle

\section{Introduction}
The rare earth iron borates, RFe$_3$(BO$_3$)$_4$ (R=rare earth, Y), belong to the trigonal system with space group R32.\cite{joubert:68} Their
structure is similar to that of the mineral CaMg$_3$(CO$_3$)$_4$ (huntite), a trigonal trapezohedral structure that is one of the five trigonal
types. This class of compounds has attracted recent attention because of a large magnetoelectric effect observed in
GdFe$_3$(BO$_3$)$_4$\cite{zvezdin:05,krotov:06,zvezdin:06,yen:06} and in NdFe$_3$(BO$_3$)$_4$.\cite{zvezdin:06b} The noncentrosymmetric
structure makes these materials also an interesting candidate for optical applications based on their good luminescent and nonlinear optical
properties.\cite{kalashnikova:04,gavriliuk:04,goldner:07} Thermodynamic and magnetic measurements have indicated a wealth of structural and
magnetic phase transitions with ordering of Fe-spins as well as the rare earth moments involved. The majority of the RFe$_3$(BO$_3$)$_4$
compounds exhibit a structural phase transition from the high-temperature structure with R32 space group (No. 155) to the low-temperature
P3$_1$21 structure (No. 152).\cite{klimin:05,fausti:06,hinatsu:03} The structure change is accompanied by a distinct anomaly of the heat
capacity.\cite{hinatsu:03,vasiliev:06} The R32 $\leftrightarrow$ P3$_1$21 transition temperature, T$_S$, scales with the ionic radius of the
rare earth and for some (R=Ho, Dy and Y) this transition takes place above room temperature. Other physical quantities, for example the
dielectric constant, exhibit distinct and sharp anomalies at T$_S$.\cite{yen:06,fausti:06}

Magnetic measurements at high temperature reveal a strong antiferromagnetic (AFM) correlation between the Fe-spins resulting in an AFM
transition at the Ne\'{e}l temperature, T$_N$, into the AFM2 phase below 40 K.\cite{hinatsu:03} Again, T$_N$ scales with the rare earth element
decreasing from 40 K (Ho) to 22 K (La). Below T$_N$, the magnetic interaction of rare earth moments with the ordered Fe-spins becomes stronger
and results in a magnetic polarization of the rare earth system and, in some rare cases, in a spin reorientation transition into the AFM1 phase
at lower temperatures that is driven by the magnetocrystalline anisotropy of the rare earth moment. The spin reorientation transition was
originally only observed in GdFe$_3$(BO$_3$)$_4$ at T$_{SR}$=9 K and its magnetic field-temperature (H-T) phase diagram was extensively
studied.\cite{balaev:03,levitin:04,pankrats:04,yen:06} However, recent neutron scattering experiments\cite{ritter:08} have found evidence for a
similar spin reorientation in HoFe$_3$(BO$_3$)$_4$ at a slightly lower temperature of 5 K. The peculiarities of the crystallographic structure,
the various possibilities for the Fe-spins to interact via direct or superexchange, and the coupling with the rare earth moments result in
complex magnetic structures and the cascade of phase transitions upon decreasing temperature, as observed in
GdFe$_3$(BO$_3$)$_4$.\cite{levitin:04} The magnetic structure has been resolved for only a few rare earth iron borates through magnetic X-ray
scattering (R=Gd)\cite{mo:08} and neutron scattering (R=Nd, Tb, and Ho) experiments.\cite{fischer:06,ritter:07,ritter:08}

Since the discovery of a large magnetoelectric effect in gadolinium iron borate,\cite{zvezdin:05} the control of the electrical polarization
through magnetic fields in GdFe$_3$(BO$_3$)$_4$ has become the focus of interest. The complex magnetoelectric H-T phase diagram was studied
through polarization, dielectric constant, magnetic, and heat capacity
measurements.\cite{balaev:03,pankrats:04,zvezdin:05,yen:06,krotov:06,zvezdin:06,kadomtseva:07} The combination of polarization and
magnetostriction experiments have proven the intimate correlation of the magnetoelastic and magnetoelectric properties of the
compound.\cite{zvezdin:05,kadomtseva:05,krotov:06} The sharp and sizable increase of the electrical polarization and the magnetostriction effect
is mainly observed at the spin reorientation phase transition between the AFM2 and AFM1 phases. The field-induced polarization (FIP) phase is
well resolved in a plateau-like structure of the dielectric constant $\varepsilon(T)$ and the complete H-T phase diagram was resolved for
different orientations of the external field H.\cite{yen:06} The magnetoelectric and magnetoelastic properties of GdFe$_3$(BO$_3$)$_4$ have been
qualitatively described by a model taking into account the symmetry of the system and the lowest order expansion of the thermodynamic potential
with respect to the magnetic exchange.\cite{kadomtseva:05}

The search for other rare earth iron borates with a significant magnetoelectric effect has been successful and NdFe$_3$(BO$_3$)$_4$ was
discovered as the second candidate showing a large field-induced polarization.\cite{zvezdin:06b,kadomtseva:07} Contrary to GdFe$_3$(BO$_3$)$_4$
no spin reorientation transition could be detected in NdFe$_3$(BO$_3$)$_4$ through magnetization, thermodynamic, and neutron scattering
experiments.\cite{fischer:06,tristan:07,popova:07b} A small maximum of the magnetic susceptibility, observed at 6 K, was attributed to
3-dimensional AFM order of the Fe$^{3+}$ and Nd$^{3+}$ sublattices.\cite{campa:97} It is remarkable that the magnetic field-induced polarization
in NdFe$_3$(BO$_3$)$_4$ exceeds the value in GdFe$_3$(BO$_3$)$_4$ by nearly two orders of magnitude. The sizable magnetoelectric coupling found
in GdFe$_3$(BO$_3$)$_4$ and NdFe$_3$(BO$_3$)$_4$ are the motivation for the search for magnetoelectric effects in other rare earth iron borates.
Ho is an interesting candidate because of a relatively large magnetic moment of the rare earth ion. Strong interactions with the Fe-spins and
possible magnetoelectric effects are therefore expected in HoFe$_3$(BO$_3$)$_4$. In fact, the magnetic order in HoFe$_3$(BO$_3$)$_4$ below T$_N$
was recently explored and it was shown to involve long range order of both, Fe-spins and Ho-moments.\cite{ritter:08} At lower temperatures,
HoFe$_3$(BO$_3$)$_4$ exhibits a spin reorientation similar to the one observed in GdFe$_3$(BO$_3$)$_4$. However, investigations of the
dielectric and magnetoelectric properties have not been reported yet.

We have therefore synthesized large single crystals of HoFe$_3$(BO$_3$)$_4$ and of the solid solution Nd$_{0.5}$Ho$_{0.5}$Fe$_3$(BO$_3$)$_4$ and
investigated their thermodynamic, magnetic, and magnetoelectric properties. We find that, even without external magnetic field,
HoFe$_3$(BO$_3$)$_4$ exhibits an electric polarization below T$_N$ which increases with decreasing temperature and suddenly drops to zero at the
spin reorientation transition. The magnetic field effect on the polarization and the H-T phase diagram is investigated.
Nd$_{0.5}$Ho$_{0.5}$Fe$_3$(BO$_3$)$_4$ shows a similar polarization effect upon decreasing temperature with an even higher value of the
polarization as compared to HoFe$_3$(BO$_3$)$_4$. In addition, Nd$_{0.5}$Ho$_{0.5}$Fe$_3$(BO$_3$)$_4$ also exhibits a large magnetoelectric
effect in external fields along the a-axis with high values of the induced polarization.

The paper is organized as follows: Section 2 provides the essential information about sample synthesis and experimental equipment used for
various measurements. The results are presented in Section 3 starting with data for HoFe$_3$(BO$_3$)$_4$ (magnetic, thermodynamic, and
dielectric data) and followed by corresponding data for Nd$_{0.5}$Ho$_{0.5}$Fe$_3$(BO$_3$)$_4$. Section 4 summarizes the main results and
conclusions.

\section{Experimental}
Single crystals of HoFe$_3$(BO$_3$)$_4$ and Nd$_{0.5}$Ho$_{0.5}$Fe$_3$(BO$_3$)$_4$ were grown using Bi$_2$Mo$_3$O$_{12}$ based flux as described
earlier.\cite{bezmaternykh:05} The crystals were cut in different sizes and shapes to fit the demands of the various measurements. The
orientation was determined by single crystal Laue X-ray diffractometry. For dielectric and polarization measurements thin plates (0.5 to 1 mm
thick) were prepared and electrical contacts were attached to two parallel faces using silver paint. For dielectric constant measurements the
high-precision capacitance bridge AH2500A (Andeen-Hagerling) was employed. The electrical polarization was extracted by integrating the
pyroelectric current measured by the K6517A electrometer (Keithley) upon variation of temperature or magnetic field. Temperature and field
control were provided by the Physical Property Measurement System (PPMS, Quantum Design). The PPMS was also employed for heat capacity
experiments. The magnetization was measured in a commercial SQUID magnetometer (MPMS, Quantum Design). Thermal expansion measurements were
conducted using a home made high-precision capacitance dilatometer.\cite{delacruz:05}

\begin{figure}
\begin{center}
\includegraphics[angle=0, width=2.5 in]{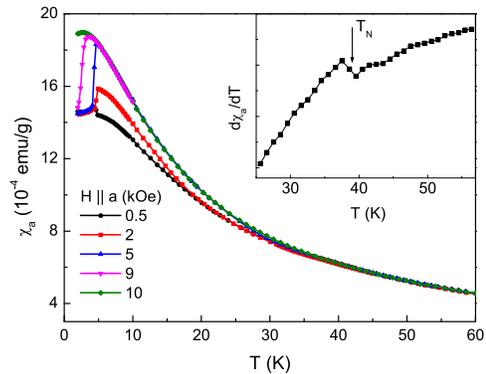}
\end{center}
\caption{(Color online) Magnetic susceptibility of HoFe$_3$(BO$_3$)$_4$ with the field oriented along the a-axis. The inset shows the
derivative, d$\chi_a$/dT, with the distinct anomaly at T$_N$.}
\end{figure}

\begin{figure}
\begin{center}
\includegraphics[angle=0, width=2.5 in]{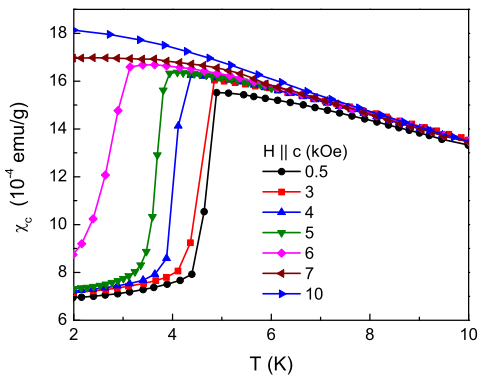}
\end{center}
\caption{(Color online) Low-temperature magnetic susceptibility of HoFe$_3$(BO$_3$)$_4$ with the field oriented along the c-axis.}
\end{figure}

\section{Results and Discussion}
\subsection{Magnetic phase diagram of HoFe$_3$(BO$_3$)$_4$}
Early investigations have shown that the structural transition from R32 to P3$_1$21 in HoFe$_3$(BO$_3$)$_4$ happens well above room temperature.
The critical temperatures, T$_S\simeq$ 420 K and T$_N\simeq$39 K, are among the highest of all rare earth iron borates.\cite{hinatsu:03} The
high magnetic ordering temperature and the easy-plane character of the Fe$^{3+}$ spin alignment was revealed also in optical measurements using
Er$^{3+}$ ions as a spectroscopic probe.\cite{stanislavchuk:07} The existence of a spin reorientation transition (T$_{SR}\simeq$5 K) in
HoFe$_3$(BO$_3$)$_4$, however, was detected only recently in neutron diffraction experiments at zero magnetic field.\cite{ritter:08}

The magnetic susceptibilities of HoFe$_3$(BO$_3$)$_4$, measured perpendicular ($\chi_a$) and parallel ($\chi_c$) to the c-axis, clearly reveal
the distinct anomalies at both magnetic transition temperatures (Figs. 1 and 2). The Ne\'{e}l transition at T$_N$=38.5 K is reflected in a
minute change of slope of the a-axis susceptibility which is clearly visible in the derivative, d$\chi_a$/dT (inset in Fig. 1). Below T$_N$ the
susceptibility shows a distinct dependence on the a-axis field, H$_a$. $\chi_a$ increases with H$_a$ below 5 kOe (Fig. 1). This behavior is
similar to the T-dependence of $\chi_a$ of GdFe$_3$(BO$_3$)$_4$ which was interpreted as an effect of the in-plane ordering of the Fe-spins
leading to a reduction of $\chi_a$ between the Ne\'{e}l and spin reorientation temperatures at zero field.\cite{yen:06} At lower temperature,
the spin reorientation transition is marked by a sharp step of $\chi_a$. This step is positive with decreasing T for low magnetic fields H$_a<$
1.2 kOe, however, it turns negative for H$_a$ between 1.2 kOe and 9 kOe. The sign change of this step-like anomaly is related to the magnetic
order induced small decrease of $\chi_a$ in the AFM2 phase since $\chi_a$ is independent of the magnetic field in the AFM1 phase (Fig. 1). The
decrease of $\chi_a$ at T$_{SR}$ is different from the corresponding increase observed in GdFe$_3$(BO$_3$)$_4$. The origin of this difference
may lie in the distinct magnetic structures of both compounds. The order of the Fe-spins is incommensurate with no long range order of the
Gd-moments above T$_{SR}$ in GdFe$_3$(BO$_3$)$_4$\cite{mo:08} but it is commensurate with the onset of Ho-moment order at T$_N$ in
HoFe$_3$(BO$_3$)$_4$.\cite{ritter:08} T$_{SR}$ quickly shifts to lower temperatures with increasing magnetic field and the AFM1 phase is
completely suppressed for H$_a>$ 10 kOe. The derived phase boundary between the AFM2 and AFM1 phases is shown in Fig. 3 (bold symbols).

The c-axis susceptibility shows no detectable anomaly at T$_N$ and very little field dependence in the AFM2 phase. At the spin reorientation
transition $\chi_c$ suddenly drops by 60 \% (Fig. 2). The resulting phase boundary is displayed in Fig. 3 (open symbols). The large decrease of
$\chi_c$ in the AFM1 phase reflects the increased correlation and stiffness of the Fe-spins after their rotation into a collinear spin structure
oriented along the c-axis that is induced by the strong coupling to the Ho moments with a large uniaxial magnetic anisotropy. The magnetic
susceptibility data reveal a minute temperature hysteresis of about 0.1 K at T$_{SR}$ indicating the first order nature of the phase transition
from the AFM2 to the AFM1 phase. The first order nature of the spin reorientation phase transition was also suggested from the step-like change
of the Ho and Fe moments derived from powder neutron scattering experiments.\cite{ritter:08} The phase diagram is further explored by isothermal
M-H measurements showing a metamagnetic transition (sudden increase of the magnetization M(H) with field) at the spin reorientation transition.
The critical fields are consistent with the phase boundaries of Fig. 3 derived from temperature dependent measurements. Similar to the $\chi$(T)
measurements the M(H) data show a small field hysteresis across the AFM2$\rightarrow$AFM1 phase boundary confirming the first order character of
the transition.

\begin{figure}
\begin{center}
\includegraphics[angle=0, width=2.5 in]{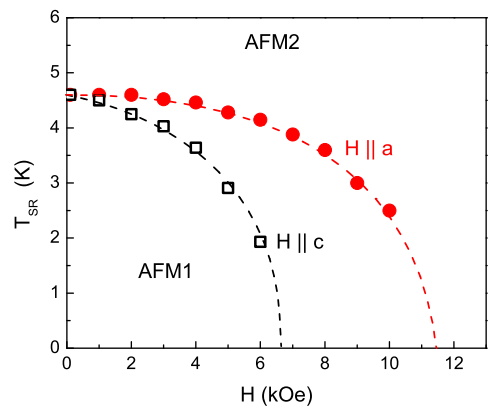}
\end{center}
\caption{(Color online) Low-temperature T-H phase diagram of HoFe$_3$(BO$_3$)$_4$ with the field aligned with the a-axis (bold symbols) and the
c-axis (open symbols). The lines serve as a guide to the eye.}
\end{figure}

The thermodynamic signature of the two magnetic phase transitions is revealed in heat capacity measurements. Figure 4 shows the heat capacity of
HoFe$_3$(BO$_3$)$_4$ together with C$_p$ of the related compounds GdFe$_3$(BO$_3$)$_4$ and ErFe$_3$(BO$_3$)$_4$ in a wide temperature range. The
inset displays the low-temperature C$_p$ of HoFe$_3$(BO$_3$)$_4$ near the spin reorientation transition on an expanded temperature scale. The
Ne\'{e}l transition is characterized by a $\lambda$-shaped pronounced peak of C$_p$(T) at 38.5 K suggesting the transition into the AFM2 phase
to be a second order phase transition. This is consistent with the steady increase of the Fe sublattice magnetization below T$_N$ observed in
neutron scattering experiments.\cite{ritter:08} However, the derived Fe magnetic moment does not follow closely the expected Brillouin function
which was attributed to the coupling of the Fe-spins with the Ho-moments leading to a simultaneous order of both magnetic subsystems at T$_N$.
The low-temperature peak of C$_p$ at T$_{SR}$ is extremely narrow as is expected at a first order transition. Below T$_N$, but above T$_{SR}$,
C$_p$/T of HoFe$_3$(BO$_3$)$_4$ shows an enhancement with respect to similar data for other rare earth iron borates, e.g. ErFe$_3$(BO$_3$)$_4$
also shown in Fig. 4. This enhancement reflects the strong coupling between the rare earth moments and the Fe spin order resulting in the spin
reorientation transition and the order of the rare earth moments. An equivalent enhancement of C$_p$/T was also observed in GdFe$_3$(BO$_3$)$_4$
(Fig. 4), the second RFe$_3$(BO$_3$)$_4$ with a spin reorientation transition.\cite{yen:06} The increased entropy associated with the enhanced
heat capacity at low T was attributed to the eight-fold degeneracy of the ground state level of the Gd$^{3+}$ ion in the corresponding crystal
field.\cite{vasiliev:06} The heat capacity data for all three compounds are compared in Fig. 4. At the lowest temperatures a Schottky-type
contribution to the heat capacity is most pronounced in ErFe$_3$(BO$_3$)$_4$ and GdFe$_3$(BO$_3$)$_4$, but barely visible in
HoFe$_3$(BO$_3$)$_4$.

\begin{figure}
\begin{center}
\includegraphics[angle=0, width=2.5 in]{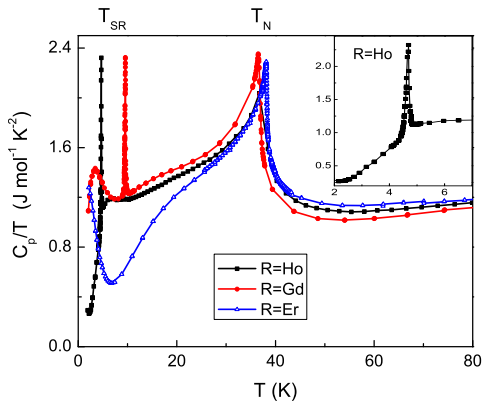}
\end{center}
\caption{(Color online) Comparison of heat capacities of HoFe$_3$(BO$_3$)$_4$, GdFe$_3$(BO$_3$)$_4$, and ErFe$_3$(BO$_3$)$_4$. The inset shows
the low-temperature heat capacity of HoFe$_3$(BO$_3$)$_4$ near T$_{SR}$.}
\end{figure}

\subsection{Magnetoelectric effect and polarization in HoFe$_3$(BO$_3$)$_4$}
\subsubsection{Dielectric constant in magnetic fields}
Dielectric anomalies below the Ne\'{e}l transition have been observed in GdFe$_3$(BO$_3$)$_4$.\cite{yen:06} A small but distinct increase of the
a-axis dielectric constant, $\varepsilon_a$, and its sudden drop at the spin reorientation transition was explained by the coupling of the
magnetic order to the lattice. While the peak-like increase of $\varepsilon_a$ in GdFe$_3$(BO$_3$)$_4$ is less than 1 \% of its value at T$_N$,
the a-axis dielectric constant of HoFe$_3$(BO$_3$)$_4$ (Fig. 5) shows a colossal enhancement of nearly 100 \% below T$_N$. This huge increase
from $\varepsilon_a$=20 to $\varepsilon_a$=37 indicates that the spin-lattice coupling in HoFe$_3$(BO$_3$)$_4$ is two orders of magnitude
stronger than the spin-lattice interaction in GdFe$_3$(BO$_3$)$_4$. With decreasing temperature $\varepsilon_a$ passes through a maximum and
drops suddenly back to its high-temperature value of 20 at the spin reorientation transition temperature, T$_{SR}$, as shown in Fig. 5.

The peak of $\varepsilon_a$ is quickly suppressed by magnetic fields oriented along the a-axis (Fig. 5a). At 4 kOe, the peak height of
$\varepsilon_a$ has dropped by 90 \% and at 10 kOe it is completely absent. This extraordinary magneto-dielectric effect shows the intricate
correlation between the magnetic order and the dielectric properties of HoFe$_3$(BO$_3$)$_4$. With the magnetic field oriented along the c-axis
(Fig. 5b) the rise of $\varepsilon_a$ below T$_N$ persists to higher field values, up to H$_c\simeq$8 kOe, beyond the critical field above which
the spin reorientation transition is suppressed (see Fig. 3). The maximum $\varepsilon_a\simeq$40 at 8 kOe is twice as large as its value near
or above T$_N$. With further increasing field, however, the maximum of $\varepsilon_a$ decreases quickly but a small enhancement below T$_N$ is
still detectable at c-axis fields as high as 40 kOe, as shown in Fig. 5b. It is remarkable that the main increase of $\varepsilon_a$ in the AFM2
phase happens well below the Ne\'{e}l temperature at about 20 K and that $\varepsilon_a$ suddenly decreases at T$_{SR}$ assuming a value in the
AFM1 phase that is close to $\varepsilon_a$ of the paramagnetic phase above T$_N$. This shows that the coupling between the magnetic order and
the lattice is most significant in the AFM2 phase at zero magnetic field but it is negligibly small in the AFM1 phase. The c-axis dielectric
constant, $\varepsilon_c$, was also measured and it shows a similar enhancement below T$_N$, however, the magnitude of the peak is much smaller
and it amounts to only 3 \% of its base value.

\begin{figure}
\begin{center}
\includegraphics[angle=0, width=2.5 in]{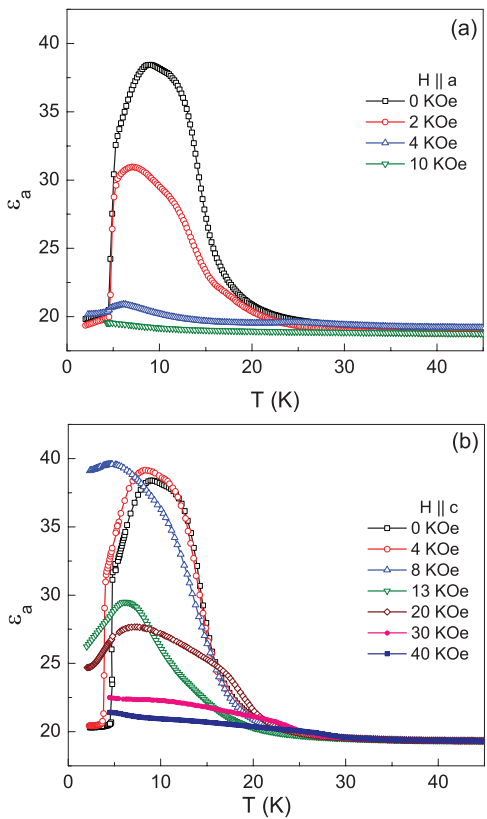}
\end{center}
\caption{(Color online) Dielectric constant, $\varepsilon_a$, of HoFe$_3$(BO$_3$)$_4$ at different magnetic fields H$\parallel$a (a) and
H$\parallel$c (b).}
\end{figure}

A similar large magneto-dielectric effect has been observed in some rare earth manganites, for example in orthorhombic HoMnO$_3$ below the
Ne\'{e}l temperature.\cite{lorenz:04b} It was later shown that the sharp increase of $\varepsilon$ was associated with the development of a
spontaneous electric polarization.\cite{lorenz:07} Therefore, we decided to search for pyroelectric and magnetoelectric effects below the
Ne\'{e}l temperature of HoFe$_3$(BO$_3$)$_4$.

\subsubsection{Polarization at zero magnetic field}
The pyroelectric current measured between two parallel electrodes attached to a sample is proportional to the change of the sample's
polarization due to a change of temperature or magnetic field. In general, a bias voltage can be applied during the measurement or it can be
used to align domains in a ferroelectric state. The integration of the pyroelectric current provides a quantitative measure of the polarization
change. However, special precautions have to be taken to ensure that the measured current actually reflects the polarization change and is free
from artifacts such as electrical transport currents or capacitive contributions due to the sample's dielectric properties. The current measured
in an experiment can be expressed by equation (1).
\begin{equation}
i=\frac{V}{R}+C\frac{dV}{dt}+\frac{dC}{dt}V+A\frac{dP}{dt}
\end{equation}
V is the applied bias voltage, R and C are the resistance and capacitance of the sample, respectively, and P is the intrinsic electrical
polarization. A is the contact area. The first term is the resistive current that can be avoided if the sample has a very high resistance or the
measurement is conducted at zero bias, V=0. The effects of charge trapping and release with changing temperature usually are not significant for
insulating materials. The second term is related to the charging current of a capacitor and will be zero if the applied voltage is constant with
time. The third contribution in (1) may cause a current if the dielectric constant (capacitance) of the sample changes significantly with
temperature (as for example $\varepsilon_a$ in Fig. 5) and a bias voltage is applied. The resulting current can be estimated from the dielectric
data and is negligible in most experiments. It can be avoided completely if no voltage bias is applied. The last term of (1) is the pyroelectric
current due to the change of the sample's polarization. In the current experiments all polarization measurements have been conducted at zero
bias voltage, V=0, which ensures that the first three terms of (1) do not cause any additional contribution to the current so that the
integration of the measured pyroelectric current yields the sample's polarization change. Measurements are commonly conducted by cooling the
sample with an applied electrical bias voltage and measuring the pyroelectric current in zero bias upon warming.

In HoFe$_3$(BO$_3$)$_4$ we found that even in zero-voltage cooling the crystal developed a pyroelectric current arising below T$_N$ at about 25
K. In passing through the spin reorientation transition the current shows a sharp negative peak indicating a sudden drop of the polarization.
Integrating the current reveals the associated change of polarization shown in Fig. 6 ((a): a-axis and (b): c-axis polarization) for cooling and
warming. Upon decreasing temperature, the polarizations measured along the a- and c-axes gradually increase in the AFM2 phase and reach a
maximum of 60 $\mu$C/m$^2$ and 90 $\mu$C/m$^2$, respectively, at T$_{SR}$. A sharp drop of P$_{a,c}$ to a small value (presumably zero, the
finite value shown in Fig. 6 is within the error limits of the measurement and integration) indicates the transition into the AFM1 phase. The
maximum values of P$_a$ and P$_c$ are comparable with the magnitude of the polarization in the ferroelectric phase of other multiferroics, for
example Ni$_3$V$_2$O$_8$ and MnWO$_4$.\cite{chaudhury:07,chaudhury:08c} However, there is a major difference in the temperature dependence of
the polarization of HoFe$_3$(BO$_3$)$_4$ in comparison with the typical behavior observed in the majority of multiferroic ferroelectric
compounds. In the latter compounds the ferroelectric polarization quickly increases below a magnetic phase transition and, in most cases, the
direction of P is determined by the direction of the bias voltage applied during electric-field cooling. In the present data for
HoFe$_3$(BO$_3$)$_4$ there is no significant change of P(T) right below the magnetic transition temperature (38.5 K) but P$_a$ and P$_c$
smoothly increase below a much lower temperature ($\simeq$ 25 K). Applying a poling voltage ($\pm$ 150 V) had no sizable effect on P(T) and the
polarization could not be reversed. Pyroelectric current data measured upon cooling with a poling voltage applied as well as upon warming after
the bias voltage was reduced to zero at 6 K are almost identical to the zero-bias data shown in Fig. 6. This indicates a strong preference of
the direction of P$_a$ and P$_c$ which may be determined by structural or magnetic domains in contrast to most multiferroic ferroelectric
systems. A clear thermal hysteresis of P(T) between 15 K and 25 K (indicated by arrows in Fig. 6) provides further evidence that the value of
the polarization in the AFM2 phase is sensitive to the magnetic domain structure.

\begin{figure}
\begin{center}
\includegraphics[angle=0, width=2.5 in]{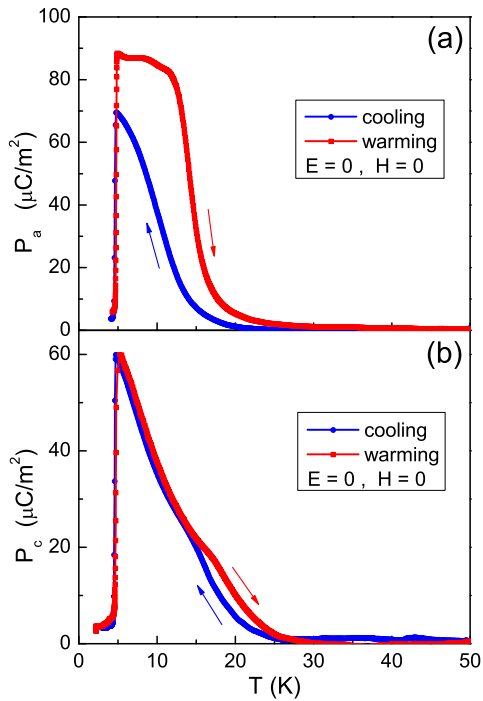}
\end{center}
\caption{(Color online) Polarization of HoFe$_3$(BO$_3$)$_4$ measured along (a): the a-axis and (b): the c-axis in zero electric and magnetic
fields. The arrows distinguish the cooling and warming data.}
\end{figure}

The most dramatic change of the dielectric constant ($\simeq$ 100 \% increase, Fig. 5) was observed along the a-axis. The maximum of the
polarization P$_a$(T) is also larger by about 50 \% than the maximum of P$_c$ (Fig. 6) and the thermal hysteresis is even more significant and
extends between T$_{SR}$ and about 30 K. To investigate the effects of electric fields (bias voltage) on the polarization P$_a$ the pyroelectric
current was measured upon cooling to 6 K (above the spin reorientation transition) with a bias of $\pm$ 200 V applied. At 6 K the external
voltage was set to zero and, after short-circuiting the contacts, the remaining polarization was measured upon warming from 6 K to 50 K. The
data shown in Fig. 7 reveal a large apparent change of the polarization due to the external voltage. At + 200 V cooling, P$_a$(+200 V) at 6 K is
enhanced by 50 \% with respect to the zero-bias P$_a$(0 V) and at - 200 V cooling, P$_a$(-200 V) is reduced to a fraction of less than 20 \%.
The solid lines in Fig. 7 represent the polarization data obtained from integrating the pyroelectric current upon cooling in + 200 V (top line),
0 V (center line), and - 200 V (bottom line). After the release of the bias voltage at 6 K, however, the polarization assumes the zero bias
value close to 70 $\mu$C/m$^2$, as indicated by the vertical dotted arrows. Upon warming P$_a$ follows closely the zero-bias temperature
dependence (dashed lines in Fig. 7). The large change of P$_a$ at 6 K upon release of the bias voltage seems to indicate a significant
tunability of the polarization in electric fields.

However, the dielectric constant $\varepsilon_a$ also increases significantly below T$_N$ (Fig. 5) and the current contribution from the third
term of equation (1) has to be taken into account in measurements conducted in an external bias electric field to determine the intrinsic
polarization values. From the dielectric measurements we determine the capacitance increase in the AFM2 phase as about 1 pF which results in an
additional charge of $\pm$ 2*10$^{-10}$ C due to a bias voltage of $\pm$ 200 V. The change of the apparent polarization at 6 K in Fig. 7
(indicated by vertical dotted arrows) amounts to 2.2*10$^{-10}$ C and -2.7*10$^{-10}$ C for a bias voltage of + 200 V and - 200 V, respectively.
After correcting for these additional contributions to the pyroelectric current, the values for P$_a$ appear to change only very little in an
electric field. Therefore, we conclude that for the a-axis pyroelectric measurement the zero-bias data (dashed lines in Fig. 7) are
representative for the intrinsic polarization, P$_a$(T). All subsequent pyroelectric measurements in magnetic fields have been conducted in zero
electric field to avoid the additional contribution from the third term in (1) to the pyroelectric current measurement. It should be noted that
this problem basically does not exist in the c-axis measurements (Fig. 6a) since the increase of $\varepsilon_c$ in the AFM2 phase is negligibly
small, only 3 \% of its value near T$_N$.

\begin{figure}
\begin{center}
\includegraphics[angle=0, width=2.5 in]{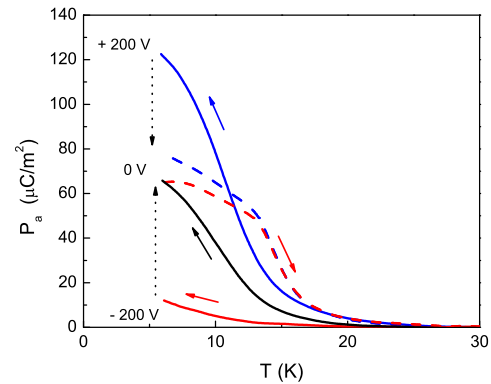}
\end{center}
\caption{(Color online) a-axis polarization of HoFe$_3$(BO$_3$)$_4$ measured upon cooling (solid lines) with applied bias voltage of +200 V, 0
V, and - 200 V. Data upon warming (dashed lines) are collected after release of the bias voltage at 6 K.}
\end{figure}

The appearance of a spontaneous polarization P(T) in the AFM2 phase and its relative insensitivity to electric fields in HoFe$_3$(BO$_3$)$_4$ is
unique among the rare earth iron borates. Changes of electrical polarization in RFe$_3$(BO$_3$)$_4$ have previously only been observed with the
application and change of an external magnetic field for R=Gd\cite{zvezdin:05} and R=Nd.\cite{zvezdin:06b} The spontaneous polarization in
HoFe$_3$(BO$_3$)$_4$ has its possible origin in the coupling to the internal magnetic field that is associated with the order of the iron spins
as well as the Ho moments. It should be noted that a polarized state is not forbidden by the crystal symmetry because the P3$_1$21 space group
is non-centrosymmetric. The magnetic order below T$_N$ and its strong coupling to the lattice results in the development of a macroscopic
electrical polarization. The preferred orientation of this polarization and the observation that it cannot be reversed by electric fields is a
result of the non-centrosymmetric structure that allows for one direction of the polarization only within a structural and/or magnetic domain.
Magnetic domains in the AFM2 phase, however, also play an important role as evidenced by the thermal hysteresis observed in the polarization
data (Fig. 6) and it lends further support to the role of the internal magnetic field/order and its mutual interaction with the polarization.
From the measured polarization values along the a- and c-axes we conclude that the intrinsic polarization direction is between both major
crystallographic orientations, at an angle of about 55 $^\circ$ with the c-axis. The origin of the intermediate orientation must be sought in
the peculiarities of the magnetic structure. According to recent neutron scattering experiments\cite{ritter:08} there exist two inequivalent
Fe-sites. In the polarized AFM2 phase the Fe(1)-spins and the Ho-moments are aligned with the a-axis, however, the spins on the Fe(2) sublattice
are tilted towards the c-axis. The estimated values of the x- and z-components of the Fe(2)-spins at 5 K are 4 $\mu_B$ and 2.5 $\mu_B$,
respectively. The angle of the Fe(2)-spins with the c-axis is therefore approximately 58 $^\circ$, close to the polarization orientation
discussed above. While it seems unlikely that this coincidence is accidental it rather demonstrates the intricate coupling of the intrinsic
polarization with the specifics of the magnetic order, supposedly of the Fe(2) sublattice.

The sharp drop of P$_a$ and P$_c$ at the spin reorientation transition indicates that the observed polarization is only allowed in the AFM2
phase with most of the magnetic moments oriented along the a-axis (some iron spins have a c-axis component, however).\cite{ritter:08} In the
low-temperature AFM1 phase all iron spins and half of the Ho moments are aligned with the c-axis and the remaining Ho moments exhibit a tilt
toward the a-axis. The particular magnetic order apparently does not couple to the lattice to induce a macroscopic polarization. The microscopic
details and the symmetry constraints have yet to be investigated. Forthcoming studies may also include the investigation of the dielectric
relaxation behavior to gather complimentary knowledge about the nature of the polarized state.\cite{schrettle:09}

\subsubsection{Magnetic field effects on the polarization}
To investigate the effects of external magnetic fields on the polarization, pyroelectric measurements have been conducted in zero electric field
upon warming from the lowest temperatures (2 K) in a constant magnetic field H applied along the a- and c-axes. Results for the electrical
polarization, P$_c$, are shown in Fig. 8. With H oriented along the a-axis P$_c$ is quickly diminished in even moderate magnetic fields (Fig.
8a). The sharp drop of P$_c$ to zero at T$_{SR}$ persists up to the critical field above which the AFM1 phase is completely suppressed. At
higher fields (10 kOe in Fig. 8a) P$_c$ remains finite to the lowest temperatures although the overall magnitude is only about 15 \% of its
maximum value at zero field. With the magnetic field applied along the c-axis (Fig. 8b) the spin reorientation transition is quickly suppressed
at low fields but the polarization retains its high values up to 8 kOe. At this field P$_c$ remains large to the lowest temperatures. Only
further increasing field reduces the magnitude of the polarization but at a lower rate as for the a-axis field. At 30 kOe the low-T polarization
is still as large as 8 $\mu$C/m$^2$. The phase boundary separating the AFM2 and AFM1 phases as derived from the polarization drop (Fig. 8) is in
perfect agreement with the phase diagrams shown in Fig. 3.

\begin{figure}
\begin{center}
\includegraphics[angle=0, width=2.5 in]{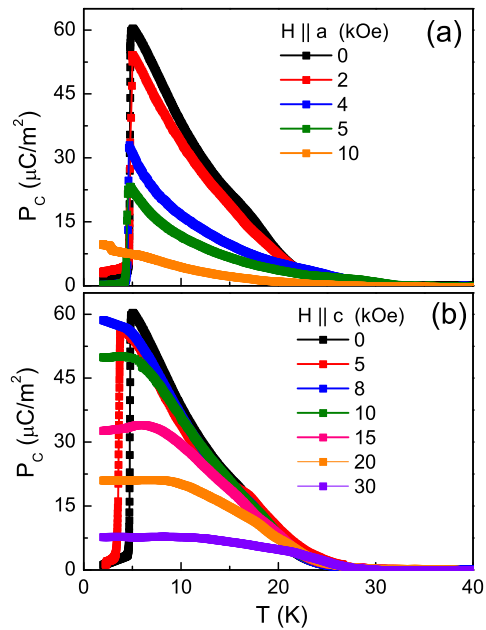}
\end{center}
\caption{(Color online) c-axis polarization of HoFe$_3$(BO$_3$)$_4$ at different magnetic fields H. (a): H $\|$ a-axis and (b): H $\|$ c-axis.}
\end{figure}

\begin{figure}
\begin{center}
\includegraphics[angle=0, width=2.5 in]{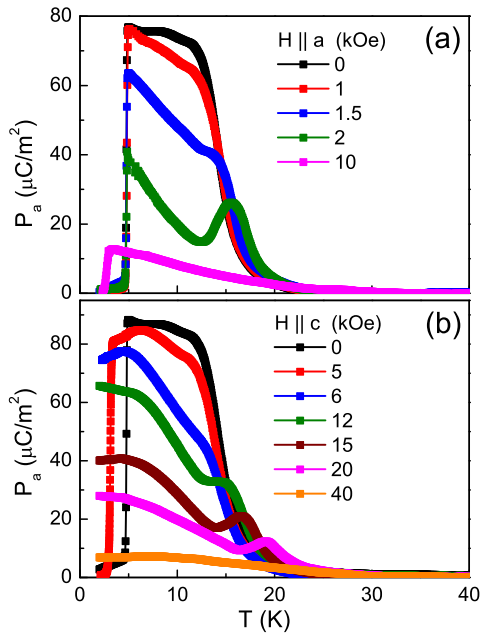}
\end{center}
\caption{(Color online) a-axis polarization of HoFe$_3$(BO$_3$)$_4$ at different magnetic fields H. (a): H $\|$ a-axis and (b): H $\|$ c-axis.}
\end{figure}

The polarization measured along the a-axis, P$_a$(T), is shown for different magnetic fields oriented along a and c in Fig. 9. P$_a$(T) reveals
a similar suppression in external magnetic fields as P$_c$(T) discussed above. The a-axis magnetic field has the strongest effect on reducing
P$_a$ and even 2 kOe are sufficient to decrease P$_a$ by 50 \%. The c-axis magnetic field suppresses the spin reorientation transition first
while maintaining a relatively high magnitude of P$_a$ up to about 10 kOe. Only higher fields H$_c$ result in a slow decrease of P$_a$ which
reaches about 10 \% of the zero-field value at H$_c$=40 kOe. The hump of P$_a$ that develops in magnetic fields between 15 K and 20 K (Fig. 9)
is associated with the temperature hysteresis in this range as shown in the zero-H polarization in Fig. 6. In magnetic fields the hysteretic
region (extending from T$_{SR}$ to about 25 K at H=0, Fig. 6a) shifts toward higher temperature. For better clarity the polarization measured
upon cooling is not included in Fig. 9. It seems surprising that the intrinsic zero-field polarization of HoFe$_3$(BO$_3$)$_4$ is quickly
suppressed by external fields whereas in the related compound NdFe$_3$(BO$_3$)$_4$ the magnetic field effect is significant and results in a
large increase of the polarization. The major differences lie in the magnetic order of rare earth and iron moments below the Ne\'{e}l
temperature. In HoFe$_3$(BO$_3$)$_4$ the spins on the Fe(1) sublattice are collinear (oriented along the a-axis) with the Ho-moments but
noncollinear with the remaining spins on the Fe(2) sublattice.\cite{ritter:08} In contrast, all Fe-spins in NdFe$_3$(BO$_3$)$_4$ are collinear
with an easy plane anisotropy and the Nd-moments are noncollinear with the Fe-spins.\cite{fischer:06} These differences in the magnetic
structures may be responsible for the different magnetoelectric properties.

\subsubsection{Thermal expansion anomalies}
The magnetoelectric coupling in multiferroic and magnetoelectric compounds are associated with ionic displacements at phase transitions that
commonly result in macroscopic distortions or lattice strain along the crystallographic
axes.\cite{delacruz:05,delacruz:06,delacruz:06c,chaudhury:07,chaudhury:08c} The relative change of the lattice parameters and the volume of
HoFe$_3$(BO$_3$)$_4$ was therefore measured and is shown in Fig. 10. The reference temperature for both axes was chosen at 55 K. The dashed
vertical lines indicate the two magnetic transitions. At T$_N$ the temperature dependence of the a- and c-axes both exhibit a minute change of
slope, as is expected for a second order magnetic phase transition. At T$_{SR}$, however, a sharp increase of c and the drop of a are
characteristic for a first order phase transition. The significant changes of the lattice parameters at T$_{SR}$ are consistent with the abrupt
disappearance of the electrical polarization and the sharp changes of the magnetic susceptibilities (Figs. 1 and 2). The large anomalies of the
lattice parameters reveal the strong magnetoelastic effect in HoFe$_3$(BO$_3$)$_4$. It is interesting that the thermal expansivities are
strongly anisotropic and the c-axis expansivity is negative over most of the temperature range shown in Fig. 10. Similar negative expansivities
have been observed, for example, in multiferroic HoMnO$_3$\cite{delacruz:05} and DyMn$_2$O$_5$.\cite{delacruz:06} It was understood as a
signature of strong magnetic correlations and spin-lattice interactions in connection with large magnetic anisotropy. The c-axis length exhibits
a maximum at about 15 K. At lower temperatures both, c and a, decrease faster with decreasing temperature approaching the instability of the
magnetic structure at the spin reorientation phase transition. The relative volume change (Fig. 10) clearly shows a rapid decrease below 15 K
leading to the magnetic instability at T$_{SR}$=4.6 K.

\begin{figure}
\begin{center}
\includegraphics[angle=0, width=2.5 in]{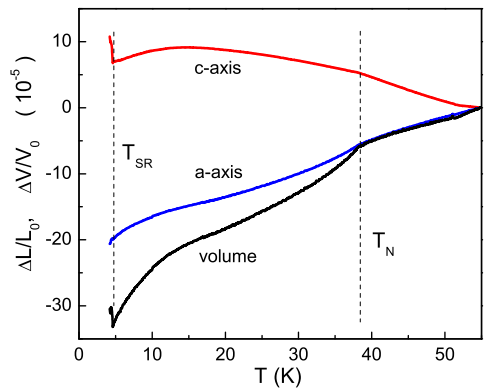}
\end{center}
\caption{(Color online) Thermal expansion of the a- and c-axes and the volume of HoFe$_3$(BO$_3$)$_4$. The two magnetic phase transitions are
marked by vertical dashed lines.}
\end{figure}

\subsection{Magnetic phase diagram, magnetoelectric effect, and polarization in Ho$_{0.5}$Nd$_{0.5}$Fe$_3$(BO$_3$)$_4$}
NdFe$_3$(BO$_3$)$_4$ does not show any evidence for a spin reorientation phase transition at low temperatures\cite{fischer:06,tristan:07} but it
exhibits a large magnetoelectric effect.\cite{zvezdin:06b} The solid solution of NdFe$_3$(BO$_3$)$_4$ and HoFe$_3$(BO$_3$)$_4$ is therefore of
interest with regard to the magnetic phase diagram and the magnetoelectric properties.

\subsubsection{Magnetic phase diagram}
The magnetic susceptibility was measured in Ho$_{0.5}$Nd$_{0.5}$Fe$_3$(BO$_3$)$_4$ for both field orientations, H$_a$ and H$_c$. $\chi_a$ and
$\chi_c$ show a sizable step-like change at T$_{SR}\simeq$9 K, clearly revealing the existence of a spin reorientation transition. While the
drop of $\chi_c$ at T$_{SR}$ is large in passing into the AFM1 phase and it amounts up to 80 \%, the change of $\chi_a$ is moderate and less
than 20 \%, similar to HoFe$_3$(BO$_3$)$_4$. However, the anomaly of $\chi_a$ changes sign at about 2 kOe. Below 2 kOe, upon decreasing
temperature, the step of $\chi_a$ is positive whereas above 2 kOe it is negative. At 2 kOe the anomaly of $\chi_a$ at T$_{SR}$ is a change of
slope with no sudden change, as shown in Fig. 11a. A similar behavior is also observed in the polarization P$_a$ in magnetic fields applied
along the a-axis (next section). It is remarkable that T$_{SR}$ of Ho$_{0.5}$Nd$_{0.5}$Fe$_3$(BO$_3$)$_4$ is twice as large as that of
HoFe$_3$(BO$_3$)$_4$. However, it is comparable with T$_{SR}$ of GdFe$_3$(BO$_3$)$_4$.\cite{yen:06} The larger value of T$_{SR}$ indicates that
Nd stabilizes the low temperature AFM1 phase and T$_{SR}$ passes through a maximum in the phase diagram of the solid solution
Ho$_{1-x}$Nd$_x$Fe$_3$(BO$_3$)$_4$. The AFM1 phase in Ho$_{0.5}$Nd$_{0.5}$Fe$_3$(BO$_3$)$_4$ is also more stable with respect to magnetic fields
as compared to HoFe$_3$(BO$_3$)$_4$. The magnetic phase diagram derived from a- and c-axis magnetization measurements is shown in Fig. 12. The
phase diagram includes data of $\chi$(T) acquired upon cooling and warming and results from isothermal magnetization measurements with
increasing and decreasing fields. In comparison to HoFe$_3$(BO$_3$)$_4$ the AFM1 phase of Ho$_{0.5}$Nd$_{0.5}$Fe$_3$(BO$_3$)$_4$ is more stable
with respect to temperature as well as magnetic field. The zero-temperature critical fields of the AFM1 $\rightarrow$ AFM2 transition are
estimated as 22 kOe and 13 kOe for H $\|$ a and H $\|$ c, respectively.

\begin{figure}
\begin{center}
\includegraphics[angle=0, width=2.5 in]{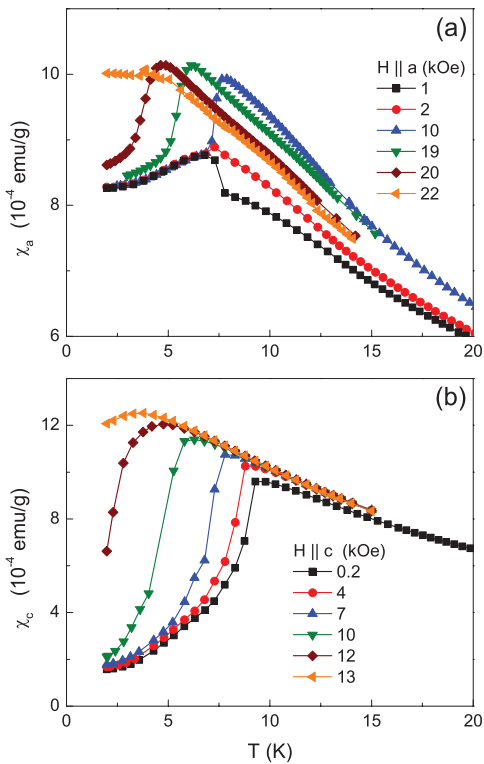}
\end{center}
\caption{(Color online) Magnetic susceptibility of Ho$_{0.5}$Nd$_{0.5}$Fe$_3$(BO$_3$)$_4$ at low temperatures in different magnetic fields. (a):
H $\|$ a-axis and (b): H $\|$ c-axis.}
\end{figure}

\begin{figure}
\begin{center}
\includegraphics[angle=0, width=2.5 in]{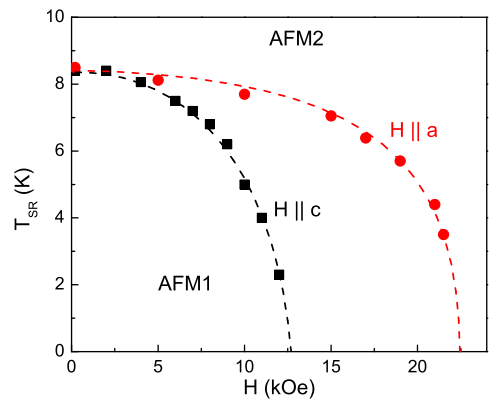}
\end{center}
\caption{(Color online) Low temperature magnetic phase diagram of Ho$_{0.5}$Nd$_{0.5}$Fe$_3$(BO$_3$)$_4$.}
\end{figure}

The thermodynamic signatures of both magnetic phase transitions, paramagnetic $\rightarrow$ AFM2 and AFM2 $\rightarrow$ AFM1, are resolved in
distinct anomalies of the heat capacity shown in Fig. 13. The $\lambda$-shaped peak at T$_N$=32 K is characteristic for a second order phase
transition. The first order spin reorientation transition is clearly marked by the sharp peak at about 9 K. While the peak at T$_{SR}$ appears
on top of a broad shoulder there is no clear maximum of C$_p$/T at lower temperatures that could indicate a Schottky type anomaly as observed
for example in NdFe$_3$(BO$_3$)$_4$\cite{tristan:07} or GdFe$_3$(BO$_3$)$_4$ (Fig. 4). The major difference to HoFe$_3$(BO$_3$)$_4$ discussed
above is the significantly lower Ne\'{e}l temperature of the iron spin order and the larger spin reorientation temperature.

\begin{figure}
\begin{center}
\includegraphics[angle=0, width=2.5 in]{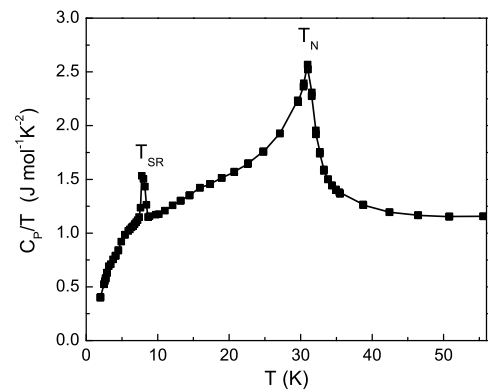}
\end{center}
\caption{Heat capacity of Ho$_{0.5}$Nd$_{0.5}$Fe$_3$(BO$_3$)$_4$.}
\end{figure}

\subsubsection{Magnetoelectric properties of Ho$_{0.5}$Nd$_{0.5}$Fe$_3$(BO$_3$)$_4$}
At zero magnetic field the spontaneous polarization measured along the a-axis, P$_a$(T), arises below 40 K in
Ho$_{0.5}$Nd$_{0.5}$Fe$_3$(BO$_3$)$_4$ and it reaches a maximum at the spin reorientation transition (Fig. 14). The maximum value of 120 $\mu
C/m^2$ exceeds the corresponding value of HoFe$_3$(BO$_3$)$_4$ by a factor of two. The transition into the AFM1 phase is accompanied by a sharp
drop of P$_a$, however, P$_a$ remains finite (about 40 $\mu C/m^2$) in the low temperature AFM1 phase. This is in distinct contrast to the
behavior in HoFe$_3$(BO$_3$)$_4$ (Fig. 9), where P$_a$ drops to zero in the AFM1 phase. With the external magnetic field applied along the
c-axis (Fig. 14b), the spin reorientation transition temperature is reduced resulting in an enhancement of the maximum of P$_a$(T) to 160 $\mu
C/m^2$ for fields up to 20 kOe. At higher magnetic fields H$_c$ the polarization decreases continuously over the whole temperature range and its
maximum value drops to 20 $\mu C/m^2$ at H$_c$=7 T.

The magnetic field dependence of P$_a$ of Ho$_{0.5}$Nd$_{0.5}$Fe$_3$(BO$_3$)$_4$ for H$\|$c is qualitatively similar to the field effect in
HoFe$_3$(BO$_3$)$_4$ as is also obvious from the isothermal polarization measurements shown in Fig. 15b. The polarization P$_a$(H$_c$) of Fig.
15 was derived from current measurements at constant temperature and increasing field. The value of P$_a$ at zero field was determined from the
T-dependent data (Fig. 14) at H$_c$=0 as the reference value. At 5 K the polarization P$_a$ is finite and nearly constant for small fields but
it increases sharply at 10 kOe because of the transition from the AFM1 to the AFM2 phase. With further increasing field P$_a$(5 K) decreases
quickly, consistent with the T-dependent data of Fig. 14b. At higher temperatures ($>$ 10 K) P$_a$ starts higher at zero field because of its
larger value in the AFM2 phase. However, the external magnetic field H$_c$ reduces P$_a$ quickly as shown in Fig. 15b. The critical temperatures
and fields of the spin reorientation phase transition determined by the sharp change of the polarization (Figs. 14b and 15b) coincide with the
c-axis magnetic phase diagram of Fig. 12 derived from magnetization measurements.

\begin{figure}
\begin{center}
\includegraphics[angle=0, width=2.5 in]{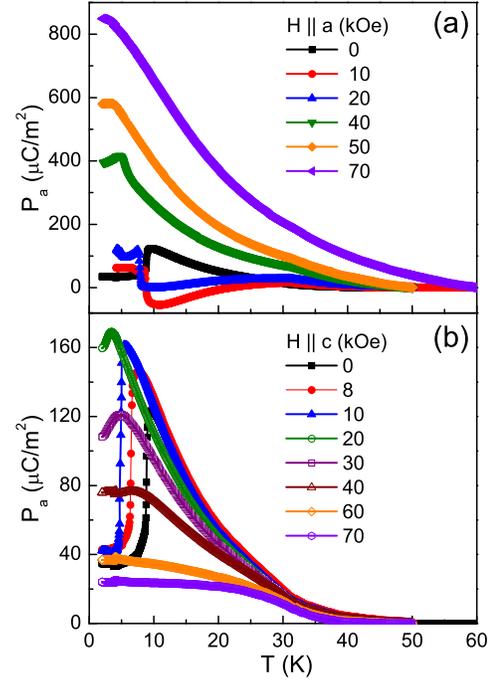}
\end{center}
\caption{Temperature dependence of P$_a$(T) of Ho$_{0.5}$Nd$_{0.5}$Fe$_3$(BO$_3$)$_4$ in different magnetic fields, (a): H $\|$ a-axis and (b):
H $\|$ c-axis.}
\end{figure}

\begin{figure}
\begin{center}
\includegraphics[angle=0, width=2.5 in]{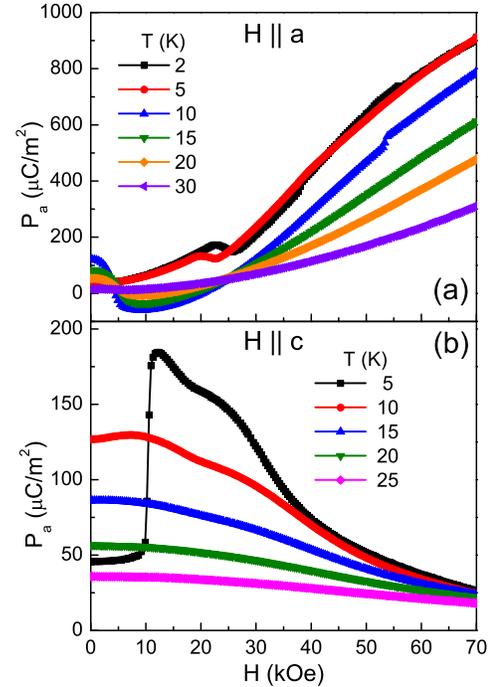}
\end{center}
\caption{Magnetic field dependence of the polarization P$_a$ at different temperatures. (a): H $\|$ a-axis and (b): H $\|$ c-axis}
\end{figure}

The effect of the c-axis magnetic field on the polarization of Ho$_{0.5}$Nd$_{0.5}$Fe$_3$(BO$_3$)$_4$ is similar to the field-induced
suppression of the spontaneous polarization of HoFe$_3$(BO$_3$)$_4$. However, the picture changes completely if the magnetic field is applied
along the hexagonal a-axis. The polarization P$_a$(T) at various H$_a$ is shown in Fig. 14a. In zero magnetic field P$_a$(T) arises below T$_N$,
exhibits the maximum at T$_{SR}$ and drops sharply to a smaller value in the AFM1 phase, as discussed above. With increasing magnetic field,
however, the positive P$_a$(T) is suppressed, it becomes negative, and passes through a minimum with decreasing temperature, as shown in Fig.
14a for H$_a$=10 kOe. The sudden decrease of P$_a$(H=0) at T$_{SR}$ is inverted in fields above 2 kOe and P$_a$ increases sharply with the
transition into the AFM1 phase. This behavior of P$_a$ at T$_{SR}$ is consistent with the anomaly of the magnetization discussed in the previous
section demonstrating the close relation of the spontaneous polarization and the magnetization.

With the magnetic field increasing further above 10 kOe the polarization P$_a$(T) starts increasing again in the whole temperature range even
above the AFM Ne\'{e}l temperature. P$_a$ is all positive for H$_a>$20 kOe and it increases to large values, P$_a>$800 $\mu$C/m$^2$, at 70 kOe
and 2 K (Fig. 14a). This high-field behavior is similar to the magnetoelectric effect observed in NdFe$_3$(BO$_3$)$_4$ and can be attributed to
a quadratic magnetoelectric coupling as discussed recently.\cite{zvezdin:06b}

The complex temperature and field dependence of P$_a$(T) shown in Fig. 14a is also reflected in the isothermal polarization data, P$_a$(H$_a$).
The results at different temperatures are displayed in Fig. 15a. At zero and small field P$_a$ is positive and determined by the spontaneous
polarization as discussed above. With increasing field the spontaneous P$_a$ is suppressed, similar to the effect of a c-axis field (Figs. 14a
and 15a) and the field effect in HoFe$_3$(BO$_3$)$_4$ (Section B). The magnetoelectric effect, since quadratic in H$_a$,\cite{zvezdin:06b} is
negligibly small. The step-like anomaly of P$_a$ near 22 kOe (2 K and 5 K data of Fig. 15a) is consistent with the temperature anomaly of P$_a$
at T$_{SR}$ (Fig. 14a). However, at higher field the magnetoelectric coupling dominates, similar to NdFe$_3$(BO$_3$)$_4$. The magnetoelectric
effect at first decreases the polarization and P$_a$(H$_a$) changes sign and becomes negative above about 4 kOe. With further increasing H$_a$
the magnetoelectric effect changes sign and results in another sign reversal of P$_a$(H$_a$) at about 20 kOe. This effect was also observed in
NdFe$_3$(BO$_3$)$_4$ and it was explained by the competition of the external magnetic field with the exchange field.\cite{zvezdin:06b} The major
difference of the polarization behavior of NdFe$_3$(BO$_3$)$_4$ and Ho$_{0.5}$Nd$_{0.5}$Fe$_3$(BO$_3$)$_4$ is the existence of a polarization
P$_a$ in zero magnetic field which is aligned with the high-field magnetoelectric polarization. The spontaneous polarization and the field
induced magnetoelectric polarization are superimposed and result in a twofold sign reversal of the total polarization upon increasing external
magnetic field between 10 K and 20 K, as shown in Fig. 15a.

We have also conducted pyroelectric measurements along the hexagonal c-axis. The detected pyroelectric current was very small and any possible
component of the polarization in zero and high magnetic fields was beyond the resolution limits of the experimental procedure.

\section{Summary and Conclusions}
We have investigated the thermodynamic, magnetic, dielectric, and magnetoelectric properties of HoFe$_3$(BO$_3$)$_4$ and
Ho$_{0.5}$Nd$_{0.5}$Fe$_3$(BO$_3$)$_4$. Both compounds show an antiferromagnetic Ne\'{e}l transition at T$_N$=38.5 K (HoFe$_3$(BO$_3$)$_4$) and
T$_N$=32 K (Ho$_{0.5}$Nd$_{0.5}$Fe$_3$(BO$_3$)$_4$) into the AFM2 phase and a spin reorientation phase transition at T$_{SR}$ into a
low-temperature AFM1 phase. The AFM1 phase is suppressed in external magnetic fields oriented along the a- and c-axes. The thermodynamic
signature of both magnetic transitions is shown in distinct anomalies of the heat capacity indicating a second order transition at T$_N$ and a
first order phase transition at T$_{SR}$. The magnetic phase diagram of both compounds is completely resolved up to 70 kOe.

The dielectric and magnetoelectric experiments on HoFe$_3$(BO$_3$)$_4$ reveal giant magneto-dielectric effect and the existence of a spontaneous
polarization below the Ne\'{e}l temperature which suddenly drops to zero at T$_{SR}$. This effect has not been observed before in rare earth
iron borates. The polarization in HoFe$_3$(BO$_3$)$_4$ has components along the a- and c-axes indicating its internal orientation in between
both crystallographic orientations. This polarization only arises in the AFM2 phase of HoFe$_3$(BO$_3$)$_4$ which shows that P is tied to the
magnetic order and the resulting internal magnetic field of this phase. In contrast to NdFe$_3$(BO$_3$)$_4$ external magnetic fields do mainly
suppress the polarization and the magnetoelectric effect is negative, dP/dH$<$0. A symmetry analysis similar to NdFe$_3$(BO$_3$)$_4$ (Ref.
\cite{zvezdin:06b}) may help to clarify the origin of the zero-field polarization of HoFe$_3$(BO$_3$)$_4$. This analysis has to take into
account the specific magnetic order of both, Fe and Ho subsystems, as known from neutron scattering data.\cite{ritter:08}

Substituting 50 \% of Ho with Nd results in a reduction of T$_N$ to 32 K and an increase of the stability range of the AFM1 phase. The
spontaneous polarization at zero magnetic field is also observed, mainly along the a-axis, with an increase of the maximum polarization (at zero
magnetic field) by 50 \% as compared to the values in HoFe$_3$(BO$_3$)$_4$. Similar to HoFe$_3$(BO$_3$)$_4$, magnetic fields along the c-axis
suppress the polarization of Ho$_{0.5}$Nd$_{0.5}$Fe$_3$(BO$_3$)$_4$. However, the a-axis magnetic field couples to the polarization and the
magnetoelectric effect induces large values of the polarization P$_a$ (up to 900 $\mu$C/m$^2$ at 70 kOe). The temperature and field dependence
of P$_a$ is very complex with several sign reversals of the polarization and step-like anomalies at the spin reorientation phase transition. The
detailed features can be understood as a superposition of the spontaneous polarization induced by internal fields and the magnetoelectric effect
due to the external magnetic field. The solid solution Ho$_{0.5}$Nd$_{0.5}$Fe$_3$(BO$_3$)$_4$ combines features observed in both parent
compounds, HoFe$_3$(BO$_3$)$_4$ and NdFe$_3$(BO$_3$)$_4$.

Further investigations of the magnetic structure, the order of Fe-spins, Ho- and Nd-moments, and the magnetoelectric interactions are necessary
to gather a basic understanding of the complex physical properties of Ho$_{0.5}$Nd$_{0.5}$Fe$_3$(BO$_3$)$_4$ and related compounds. The present
investigation may be extended to include the solid solution of other members of the rare earth iron borate compounds. One interesting candidate
could be the solid solution of Ho$_{1-x}$Gd$_{x}$Fe$_3$(BO$_3$)$_4$. GdFe$_3$(BO$_3$)$_4$ is the only second compound in the rare earth iron
borate system to show a significant magnetoelectric effect and, therefore, the magnetoelectric properties of
Ho$_{1-x}$Gd$_{x}$Fe$_3$(BO$_3$)$_4$ should be interesting too. A thorough investigation of Ho$_{1-x}$Gd$_{x}$Fe$_3$(BO$_3$)$_4$ could even help
to arrive at a deeper and more comprehensive understanding of the current results on Ho$_{0.5}$Nd$_{0.5}$Fe$_3$(BO$_3$)$_4$.

\begin{acknowledgments}
This work is supported in part by NSF Grant No. DMR-9804325, the
T.L.L. Temple Foundation, the J. J. and R. Moores Endowment, and the
State of Texas through the TCSUH and at LBNL by the DoE.
\end{acknowledgments}

\bibliographystyle{phpf}

\end{document}